\documentclass[iop]{emulateapj}

\shorttitle{Substructure around the Hercules dSph}
\shortauthors{Roderick et. al.}

\usepackage{amsmath,amssymb}

\begin{document}

\title{Stellar Substructures around the Hercules Dwarf Spheroidal Galaxy}
    
    \author{T. A. Roderick \altaffilmark{1}}
    \author{H. Jerjen\altaffilmark{1}}
    \author{A. D. Mackey\altaffilmark{1}}
    \author{G. S. Da Costa\altaffilmark{1}}
\affil{Research School of Astronomy and Astrophysics, Australian National University,\\ Canberra, ACT 2611, Australia}
\email{Tammy.Roderick@anu.edu.au}

\begin{abstract}
We present deep $g,i$-band DECam stellar photometry of the Hercules Milky Way satellite galaxy, and its surrounding field, out to a radial distance of 5.4 times the tidal radius. We have identified nine extended stellar substructures associated with the dwarf; preferentially distributed along the major axis of the galaxy.  Two significant over-densities lie outside the 95\% confidence band for the likely orbital path of the galaxy and appear to be free-floating tidal debris.  We estimate the luminosity of the new stellar substructures, and find that approximately the same amount of stellar flux is lying in these extended structures as inside the main body of Hercules.  We also analyse the distribution of candidate blue-horizontal-branch stars and find agreement with the alignment of the substructures at a confidence level greater than 98\%.  Our analysis provides a quantitative demonstration that Hercules is a strongly tidally disrupted system, with noticeable stellar features at least 1.9 kpc away from the galaxy.

\end{abstract}

\keywords{galaxies: dwarf (galaxies:) Local Group}

\section{Introduction}
\label{sec:intro}
Ultra-faint Milky Way (MW) satellite galaxies are the most dark matter dominated stellar systems in the Universe \citep{1998ARA&A..36..435M, 2007ApJ...670..313S, McConnachie:2012fh}.  The high mass-to-light ratios seen in these pressure-dominated systems are determined from their velocity dispersion, which assumes that the underlying stellar populations are in dynamic-equilibrium.  However, satellite galaxies interacting with their hosts can undergo significant tidal stirring \citep{2012ApJ...751...61L}, leaving kinematic samples potentially contaminated by unbound stars \citep{2007MNRAS.378..353K}.  In the case where a galaxy is being tidally disrupted, the mass-to-light ratio can be overestimated, and this may not be apparent in the galaxy's dispersion profile \citep{2008ApJ...673..226P}. It is also possible for a galaxy that has undergone tidal disruption to retain its spherical shape \citep{2008ApJ...679..346M}.  

The dynamical state of a satellite galaxy is thus a highly relevant question.  For that reason it is important to search for signs of tidal disruption not only in the central region of these galaxies, but in the vicinity around them where tidal debris may be present.    The MW satellites provide an excellent opportunity for this type of investigation, as they are close enough to be resolved into individual stars and can be studied in great detail \citep[see][for a discussion and census of local satellites]{2009ARA&A..47..371T, McConnachie:2012fh,2012PASA...29..383J}. 

Comprehensive studies of the MW satellite galaxies have led to numerous investigations of their role in galaxy formation.  In the $\Lambda$CDM cosmological paradigm, galaxies form hierarchically through merger and accretion of smaller structures \citep{1984Natur.311..517B, 2005MNRAS.359.1029V, Giocoli:2009vw}.  \citet{1999ApJ...524L..19M} demonstrated that dark matter sub-structure occurs on galactic scales, resulting in galaxy halos appearing as scaled versions of galaxy clusters.  Mergers and accretion are a feature of galaxy clusters, therefore the outskirts of larger galaxies should also show signs of merger and accretion events \citep{2007ApJ...667..859D}.  A rather striking example in this context is Andromeda (M31), which possesses numerous satellite systems and copious substructure in its stellar halo \citep{2009Natur.461...66M, 2014ApJ...780..128I}. Substructures are also observed in our own Milky Way in the form of stellar streams \citep[e.g.][]{2002AAS...201.6502N, 2007ApJ...658..337B, 2006ApJ...651L..29G, 2010ApJ...711...32N, 2013ApJ...776...26S, 2014ApJ...790L..10G}, with perhaps the quintessential example being the Sagittarius dwarf \citep{1994Natur.370..194I}, and its great tidal tails strewn across the sky \citep{2003ApJ...599.1082M,2013AJ....145..163N}.  

Contrary to the predictions of $\Lambda$CDM, \citet{Kroupa:2005iy} demonstrated that the distribution of the MW satellite population is inconsistent with that of a dwarf galaxy population drawn from cosmological sub-structure.  Further investigation of the satellite populations of both the MW and M31 has found, in both cases, that the satellites have a disc-like distribution with high inclination to the plane of the host galaxy \citep{2005nfcd.conf..259M, Metz:2007ch, 2009MNRAS.394.2223M, 2013MNRAS.435.1928P, 2013Natur.493...62I, 2013ApJ...766..120C}.  This has led to the alternative picture that a significant fraction of the currently known MW satellite galaxies is of tidal origin \citep{2010A&A...523A..32K, 2014MNRAS.442.2362P}; forming from a major collision of the MW and another galaxy.  This theory accounts for many of the observed features of the dwarf galaxy population, including the high mass-to-light ratios (explained as systems driven out of equilibrium \citep{2014MNRAS.442.2419Y}).

Learning whether or not the dwarf population is largely comprised of systems driven out of equilibrium, or shows signs more indicative of hierarchical build-up, may be a key factor in determining the origin of the population as a whole.  Most of the research into the MW dwarf galaxy population to date has been restricted to the main stellar body of each dwarf.  However, it is in the outer regions where we expect to see more subtle signs of tidal disruption.  The obvious example is Sagittarius \citep{1994Natur.370..194I, 2013AJ....145..163N}, however, Ursa Minor \citep{2003AJ....125.1352P}, Ursa Major II \citep{2007ApJ...670..313S}, Carina \citep{Battaglia:2012it,2014MNRAS.444.3139M}, and Fornax \citep{2004AJ....127..832C} all display stellar substructure in their outer regions, suggesting that they are not dynamically pristine stellar systems.

The need for wide-field studies of the dwarf galaxy population is a recurring theme.  Given recent advances in large-mosaic CCD cameras, a logical next step is to choose a target and explore its surrounding area in detail.  The Hercules dwarf spheroidal galaxy (Table \ref{table:Herc}) detected in the Sloan Digital Sky Survey \citep{2007ApJ...654..897B} is, in this context, particularly interesting.  It is the most elongated of the MW dwarfs \citep[see][]{2007ApJ...668L..43C, 2009ApJ...704..898S, 2012MNRAS.425L.101D}, which has led to suggestions of tidal disruption and interaction with the MW. \citet{2007ApJ...668L..43C} noted that, given the large heliocentric distance of Hercules at 132\,kpc, in order for it to be tidally disrupted, it must be on a highly elliptical orbit about the MW; consistent with the best model value for the perigalactic distance of $r_p=18.5$\,kpc \citep{2009ApJ...706L.150A}. \citet{2009ApJ...704..898S} suggested that Hercules may be embedded in a stellar stream, finding a stellar extension to the northwest of Hercules, and several associated stellar over-densities along the major axis out to $\sim35\arcmin$ (1.4\,kpc).  \citet{2010ApJ...721.1333M} further developed the idea that Hercules is part of a stream, and suggested that a progenitor on its orbit may have been taken close enough to the MW to induce disruption from a bound dwarf galaxy into a stellar stream.  They present a viable scenario in which this is the case, where the orbital path of Hercules is aligned with its direction of elongation. The Hercules stellar population is old ($>$12\,Gyr) and metal poor ([Fe/H]=-2.41) \citep{2007ApJ...668L..43C, 2008ApJ...685L..43K, 2011A&A...525A.153A, 2014A&A...570A..61F}, and displays a spread in [Fe/H] of more than 0.5 dex \citep{2011A&A...525A.153A,2011ApJ...727...78K}. It has also been shown to have a high level of $\alpha$-enhancement \citep{2008ApJ...688L..13K, 2011A&A...525A.153A, 2012AIPC.1480..190K}, suggesting a large chemical enrichment contribution from TypeII supernovae progenitors.  More recently, \citet{Koch:2013eu} reported a significant deficiency in neutron-capture elements and suggested that the chemical enrichment of Hercules was governed by massive stars, coupled with a low star formation efficiency.  All of this indicates Hercules has a rather colourful star formation history. 

\begin{table}
\begin{center}
\caption{Fundamental Parameters of Hercules}
\begin{tabular}{llc}
\tableline\tableline
Parameter & Value & Ref.\tablenotemark{a}\\
\tableline
Morphological type & dSph& (1)\\
RA (J2000) & 16:31:02& (1)\\
DEC (J2000) & +12:47:30& (1)\\
\textit{l} & $28.7^\circ$& (1)\\
\textit{b} & $36.9^\circ$& (1)\\
$D_\odot$ & $132\pm12$\,kpc& (2)\\
$M_V$ & $-6.6\pm0.4$& (3)\\
$(m-M)_0$ & $20.6 \pm 0.2$& (2)\\
$v_\odot$ & $45.2\pm1.1$km s$^{-1}$& (4)\\
$\sigma$ & $3.7\pm0.9$ km s$^{-1}$& (4)\\
$\epsilon$ & $0.63\pm0.02$& \\
$\theta$ & $113.8\pm0.6^\circ$& \\
$r_h$ & 330\,pc& (3)\\
$[Fe/H]$ &  $-2.41\pm0.04$& (5)\\
Stellar mass (dynamic) & $2.6\times 10^6 M_\odot$& (6)\\
Mean age & $>$12 Gyr& (6)\\
Mass-to-Light ratio & $103 \substack{+83 \\ -48}$& (7)\\
$r_{tidal}$ (theoretical) & 485\,pc& (7)\\
\tableline\tableline
\label{table:Herc}
\end{tabular}
\tablenotetext{1}{\textbf{References:} (1) \citet{2007ApJ...654..897B}, \\(2) \citet{2007ApJ...668L..43C}, (3) \citep{2008ApJ...684.1075M},\\ (4) \citet{2009A&A...506.1147A}, (5) \citet{2012PASA...29..383J}, \\(6) \citet{2009ApJ...704..898S}, (7) \citet{2009ApJ...706L.150A}}
\end{center}
\end{table}

\begin{figure}
\includegraphics[width=8cm]{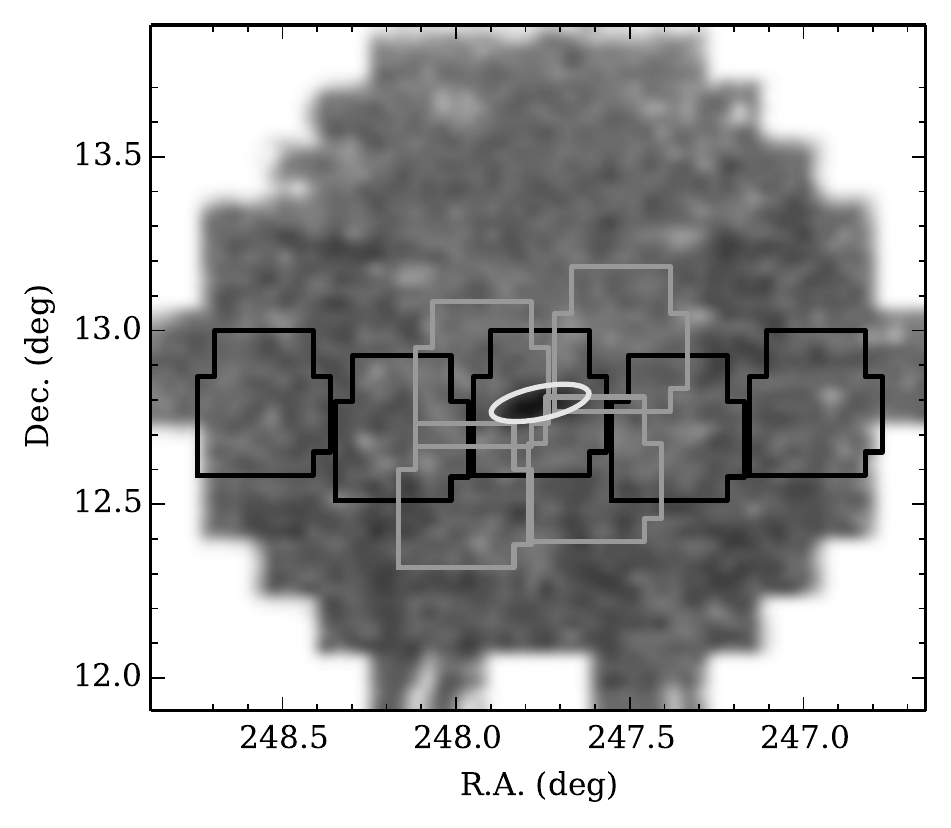}
\caption{Field-of-view of our DECam observation (showing a smoothed representation of our complete catalogue from Section \ref{sec:obs}).  The ellipse in the centre delineates the orientation and half light radius of Hercules \citep{McConnachie:2012fh}. Outlines for the five fields from \citet{2009ApJ...704..898S} are superimposed in black, while the four fields from \citet{2014A&A...570A..61F} are shown as a dashed outline.}
\label{fig:decam_fov}
\end{figure}

With the exception of  \citet{2009ApJ...704..898S}, and more recently \citet{2014A&A...570A..61F}, all observations of Hercules have focussed on its main stellar body \citep[e.g.][]{2007ApJ...668L..43C, 2009A&A...506.1147A, 2011A&A...525A.153A}.  \citet{2009ApJ...704..898S} made observations along the semi-major axis, with five 23$\arcmin \times$ 23$\arcmin$ fields, and  \citet{2014A&A...570A..61F} made similar observations with a $2\times2$ arrangement of  23$\arcmin \times$ 23$\arcmin$ fields centred on Hercules.  In this paper, we make use of the large field-of-view afforded by the Dark Energy Camera (DECam) on the 4m Blanco telescope at Cerro Tololo in Chile. We present wide-field observations (see Figure \ref{fig:decam_fov}) in order to determine the true nature of Hercules' elongated shape; including how far it extends into its surrounds, both parallel and perpendicular to the major axis.  In Section \ref{sec:obs}, we describe the observations and data reduction process, including photometry and star/galaxy separation.  We also discuss photometric calibration and completeness.  In section \ref{sec:discrimination}, we describe the process of discriminating between Hercules and MW foreground stars via model isochrone selection.  Section \ref{sec:analysis} contains our analysis of the shape of Hercules and the identification of statistically significant over-densities, while in Section \ref{sec:results} we investigate the luminosity of these over-densities and perform a further test of significance using blue-horizontal-branch star candidates as tracers.  In section \ref{sec:summary}, we discuss these results in relation to the tidal disruption of Hercules, and summarise our findings.

\section{Observations and Data Reduction}
\label{sec:obs}
Observations were carried out over two nights on 2013 July 12 and July 13, using DECam at the CTIO 4m Blanco telescope, as part of observing proposal 2013A-0617 (PI: D. Mackey). DECam is comprised of a hexagonally shaped mosaic of 62 2K$\times$4K CCDs, each with a pixel scale of $0\farcs27$/pix, providing a total field-of-view of 3 square degrees. 

The data set is composed of a single DECam pointing, encompassing more than five times the estimated tidal radius of Hercules  \citep[485\,pc or $12\farcm6$,][]{2007ApJ...668L..43C, 2009ApJ...706L.150A}, with 2 sets of 4$\times$ 900s exposures in \textit{g}, and 1 set of 11$\times$ 600s exposures in \textit{i}. Relatively poorer seeing in the initial set of \textit{g}-band images resulted in a second set of exposures being taken to obtain appropriate photometric depth, thus giving total integration times of 7200s and 6600s in \textit{g} and \textit{i} respectively.  A box shape dither pattern was used for exposures in both filters.  This was to fill in the inter-chip gaps on the focal plane, and to avoid stars repeatedly falling on the same pixels (reducing the effect of bad pixel columns and CCD edge pixels).

The images were reduced using the DECam community pipeline\footnote{http://www.ctio.noao.edu/noao/content/Dark-Energy-Camera-DECam} \citep{2014ASPC..485..379V}.  This process included the application of a WCS solution to each image, the subtraction of sky-background, and the co-addition of images into a single image stack for each filter.  The final product is a multi-extension FITS file, containing the stacked image-mosaic sliced into 9 separate tiles; one tile per FITS extension. The combined images for the Hercules data set have an average seeing across the field-of-view of $1\farcs49$ for \textit{g} and $1\farcs09$ for \textit{i}. 

The full-frame subtraction of the sky background from the hexagonal DECam images resulted in the corners of the final stacked FITS image containing no science data, but having non-zero pixel values. During preliminary photometry, this was found to cause problems with the background mapping for the corner tiles, where as much as 50\% of the tile contained non-science pixels.  In order to overcome this issue, WeightWatcher\footnote{http://www.astromatic.net/software/weightwatcher} \citep{2008ASPC..394..619M} was used, in conjunction with the weight maps produced by the  community pipeline, to mask out the non-science pixels.  By using these weight maps, the background map was weighted to reflect the presence of bright stars and the different levels of exposure created by dithering the CCD mosaic, producing an optimised background map for conducting photometry.

 \subsection{Photometry}
 \label{sec:photom}
Aperture photometry was performed on each of the nine tiles of the \textit{g} and \textit{i} images using the program Source Extractor\footnote{http://www.astromatic.net/software/sextractor} \citep{1996A&AS..117..393B}.  Each of the corresponding masked weight maps from WeightWatcher were used by Source Extractor to produce the background map before source detection and photometry were carried out. The extraction process was run in two passes for each tile.  A shallow first pass determined the average FWHM ($\overline{F}$) value for point sources in the image, where point sources were defined to be bright, circular detections with a Source Extractor `internal flag' of 0 indicating isolation from near neighbours or image edges.  This information served as input for the second, deeper pass, where aperture photometry was measured within $1\times$$\overline{F}$ and $2\times$$\overline{F}$.  Using this method separately on each of the nine tiles also ensured that the aperture size was allowed to reflect variations in the point spread function (PSF) across the large field-of-view.  

\begin{figure}
\centering
\includegraphics[width=8cm]{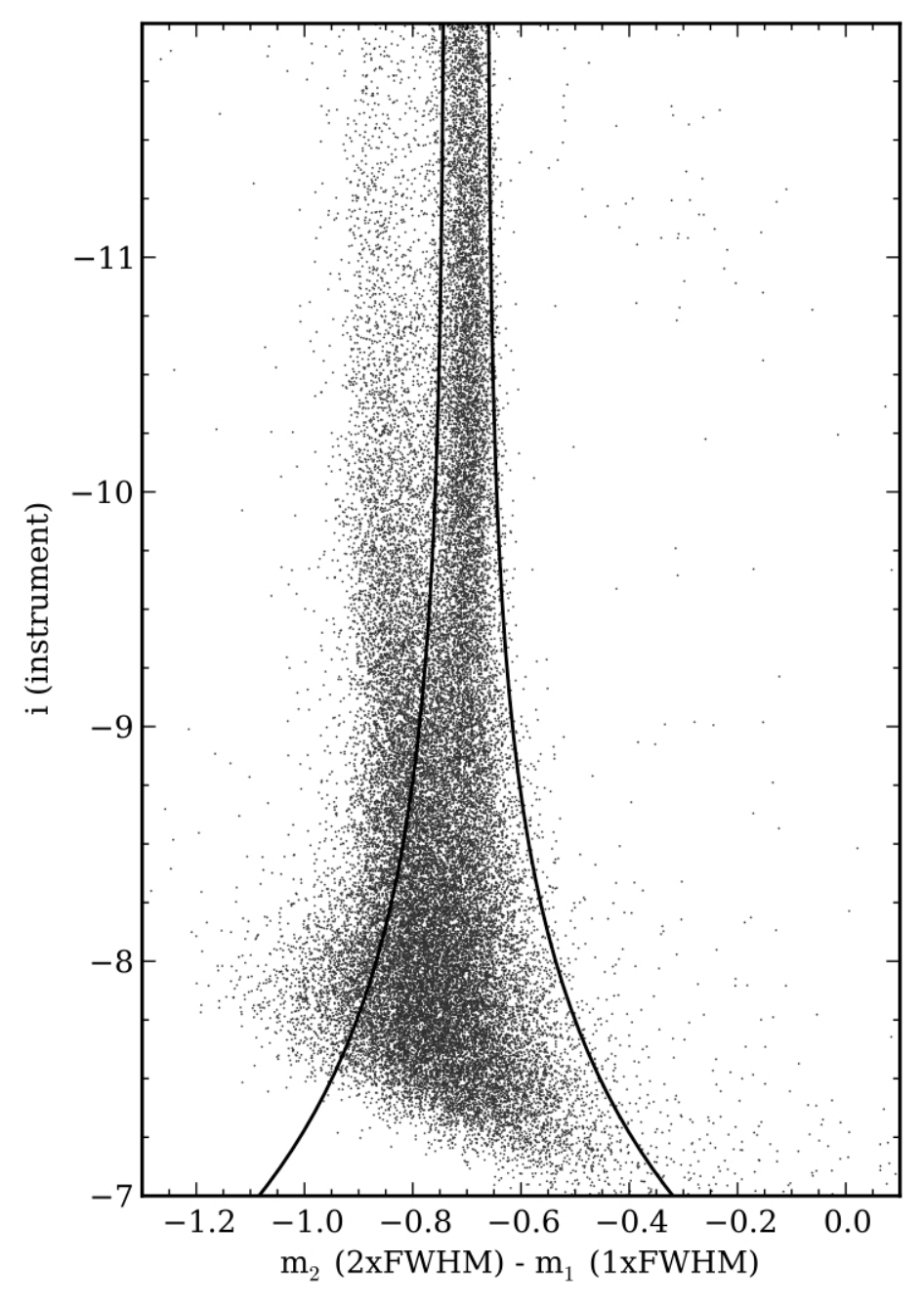}
\caption{Example of the star/galaxy separation method used, where objects are distinguished by their light profile. $m_1$ corresponds to flux measured inside an aperture of $1\times$ $\overline{F}$ and similarly, $m_2$ an aperture of $2\times$ $\overline{F}$.  Background galaxies are seen as a plume to the left of the vertical column of stellar objects. }
\label{fig:apertures}
\end{figure}

Spurious and non-stellar detections were eliminated from the Source Extractor catalogues by discarding objects whose light profile appeared significantly different to the point-like sources in the catalogue.  Different types of objects were identified by examining the difference in instrument magnitude, $m_2-m_1$ (where $m_1$ and $m_2$ correspond to aperture sizes of  $1\times$ $\overline{F}$ and  $2\times$ $\overline{F}$ respectively).  Stars with a well defined PSF populate a narrow range of negative $m_2-m_1$ values. Background galaxies however, are more extended and have a larger relative flux encompassed in the $m_2$ aperture than the $m_1$ aperture.  Objects with $m_2-m_1 \geq 0$ represent cosmic rays, or other spurious objects which were instantly discarded. Figure  \ref{fig:apertures} shows $m_2-m_1$ against instrumental magnitude for a single image tile in the \textit{i}-band. Note that saturation occurs at an instrumental magnitude of $\simeq-12$, and therefore detections brighter that this were excluded from analysis. The stars are the main vertical distribution of points (centred at $m_2-m_1\simeq-0.68$),  flaring out at the faint end.  Extended sources appear as a plume to the left of this population.  The boundaries of the main stellar distribution were modelled with an exponential function to create a region of acceptability.  Objects that fell outside this region were excluded from the catalogue.  Our own star/galaxy separation method was accompanied by a cut based on the Source Extractor star/galaxy flag.  A loose cut was made with objects accepted as stars if the flag was $>0.35$.  This method of combining the light profile and Source Extractor flag was adopted in order to retain more stellar objects at fainter magnitudes. The Source Extractor classification flag becomes less reliable at faint magnitudes due to the ambiguity in determining the shape of objects at this level; experimentation revealed that adopting this combined methodology gained at least a magnitude in depth with relatively little cost or contamination.

The star/galaxy separation was performed only on the \textit{i} band image, due to the significantly poorer seeing in the \textit{g} band.  The results of this process were applied to the \textit{g} band during the following cross matching step.

The last step in creating the final catalogue of stellar sources was to cross reference the WCS solution coordinates, applied during the community pipeline processing, between the \textit{g} and \textit{i} catalogues.  The cross referencing was carried out using the Stilts command line package \citep{2011ascl.soft05001T}.  Only stars detected in both filters, and passed through the \textit{i} band star/galaxy separation, were included.

\begin{figure}
\centering
\includegraphics[width=8cm]{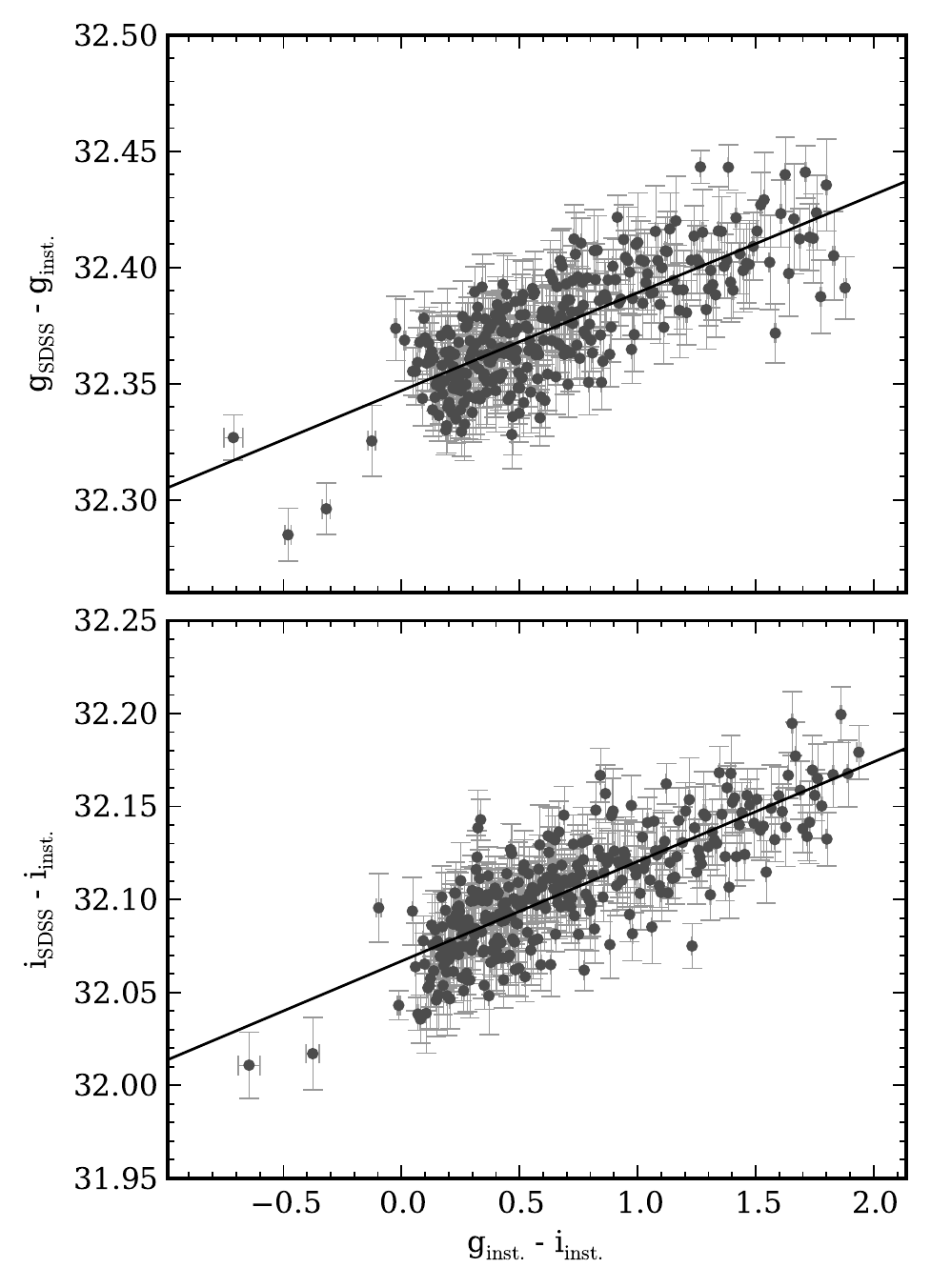}
\caption{Comparison  of the Hercules catalogue with the SDSS photometry, used to determine the offset and colour terms for each filter. These values in \textit{g} and \textit{i} respectively are: $Z_g=32.347$,  $c_g=0.042$,  $Z_i=32.067$,  $c_i=0.054$. Each point and error bar represents the mean and standard deviation of 10 stars, first sorted by colour and then grouped into tens.} 
\label{fig:calibration}
\end{figure}

Once the complete catalogue of 87,000 stellar sources was established, it was calibrated onto the standard SDSS photometric system. This was achieved by first matching objects from the Hercules catalogue to stars from SDSS data release 10 \citep[SDSS DR10:][]{2014ApJS..211...17A}. The photometric zero points (Z), and colour terms (c), where colour is defined as $g_{inst}-i_{inst}$,  were determined by comparing the SDSS magnitude to the instrumental magnitude of the $2\times$ FWHM aperture for each star.  Figure \ref{fig:calibration} shows this comparison for each filter (each point on the plot represents the mean of 10 stars, first sorted by colour and then grouped into tens).  The following values were determined for \textit{g} and \textit{i}, using a least-squares fit:

\begin{displaymath}
Z_g=32.347\pm0.002, c_g=0.042\pm0.002\\
\end{displaymath}
\begin{displaymath}
Z_i=32.067\pm0.002, c_i=0.054\pm0.002\\
\end{displaymath}

\begin{figure}
\centering
\includegraphics[width=8cm]{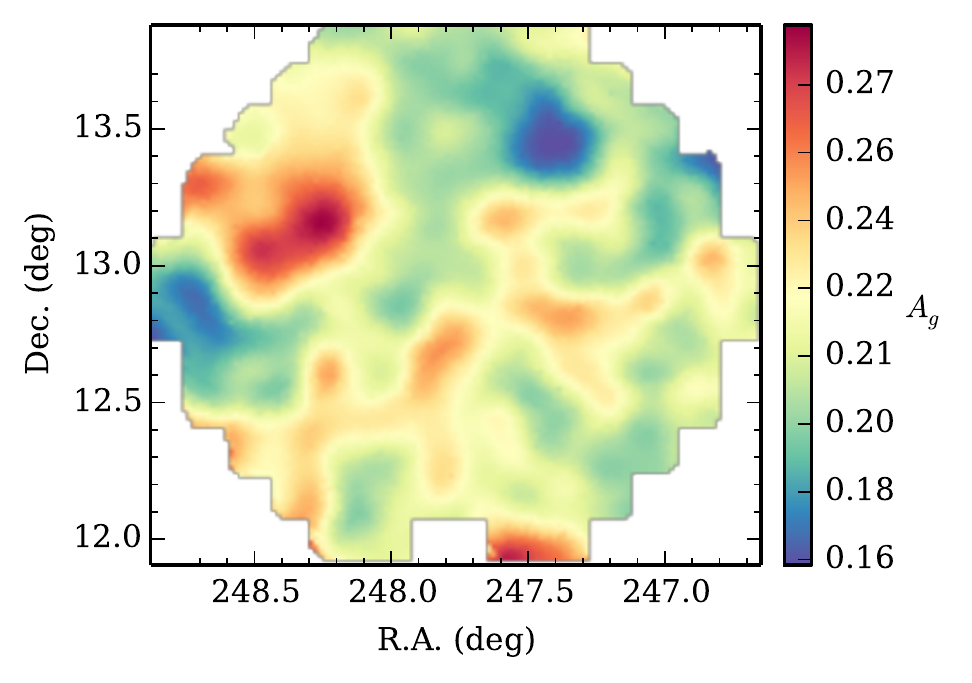}
\caption{Map of Galactic extinction ($A_g$) used for each star in the Hercules catalogue.  Correction values used are those from the SDSS photometry, based on \citet{1998ApJ...500..525S}.}
\label{fig:extinction}
\end{figure}

The large field-of-view afforded by DECam begets a significant variation in Galactic extinction across the field. It was, therefore, deemed inappropriate to use a single correction value for the whole Hercules catalogue.  Instead, the extinction correction values ($A_g$ and $A_i$) from SDSS DR10, based on \citet{1998ApJ...500..525S}, were applied to the calibrated Hercules photometry.  The specific coefficients used were $A_g=3.793 E(B-V)$ and $A_i=2.086 E(B-V)$ \citep{2002AJ....123..485S}. The values were applied by matching each star in the catalogue to its nearest SDSS neighbour, and using the corresponding extinction value. Figure \ref{fig:extinction} illustrates the extinction variation $0.165 < A_g < 0.280$ (a variation of approximately 5\% in apparent luminosity) across the field, and corresponds to the extinction maps of \citet{1998ApJ...500..525S} in the direction of Hercules.  The final, fully calibrated and extinction corrected catalogue was used as the reference for the following analysis.
\\
\\

\subsection{Completeness}
\label{sec:completeness}
\begin{figure}
\centering
\includegraphics[width=8cm]{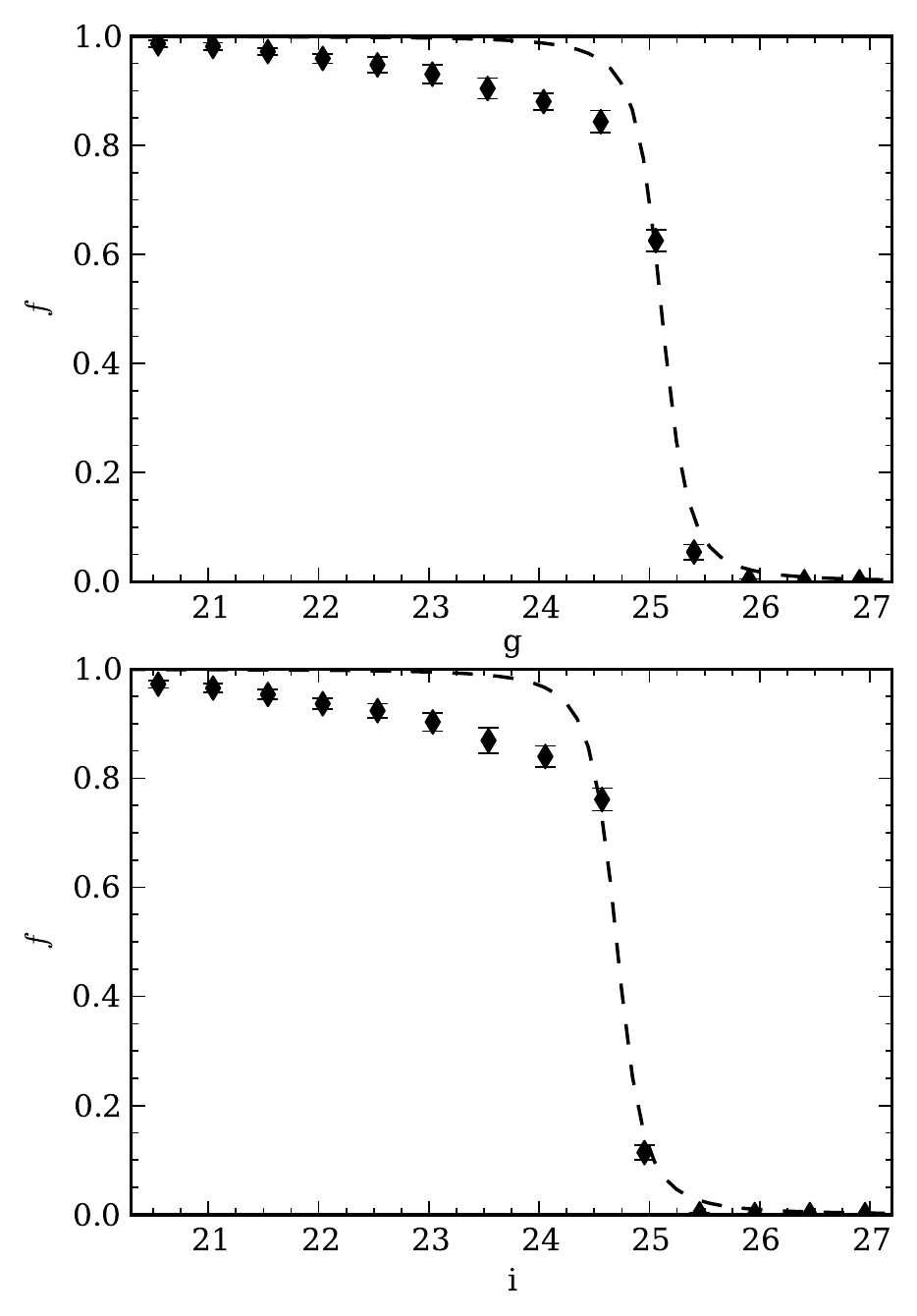}
\caption{Black points show recovered artificial stars for each 0.5 magnitude bin averaged across the field (in $g$ and $i$).  The dashed lines show the best-fitting relation of the form described by \citet{1995AJ....109.1044F}, which was used to determine the 50\% completeness levels.}
\label{fig:completeness}
\end{figure}

Artificial star tests were carried out in order to estimate the photometric depth of the data set, and determine the photometric accuracy.  The IRAF \textit{addstar} task was used to add stars of random R.A. and Dec. to each image tile; 40 stars each in 0.5 mag intervals between 20.5 and 27.0 mag.  The previously described photometry pipeline was then run on each tile to recover the artificial stars.  This process was repeated 30 times, to determine the mean recovery rate for each image tile.  The completeness level of the whole image was estimated by taking the average over the nine tiles.  The 50\% completeness limit was ascertained by fitting the interpolation function described by \citet{1995AJ....109.1044F} to the distribution of recovered artificial stars (shown in Figure \ref{fig:completeness}), and determined to be 25.11 in \textit{g} and 24.70 in \textit{i}, respectively. 

The photometric accuracy was determined by measuring the variation in recovered magnitudes for each input magnitude.  An exponential was fit to these measurements and used as a function for determining the photometric uncertainty for individual stars in the catalogue.  This procedure was completed separately for each photometric band.

\section{Foreground Discrimination}
\label{sec:discrimination}
In order to detect stellar over-densities related to the Hercules dwarf galaxy, it is essential to disentangle its stellar population from that of the MW foreground.  Selecting stars based on their position in the colour-magnitude diagram (CMD) is a useful way of performing such discrimination, and a model isochrone provides an excellent aid.

\begin{figure*}
\centering
\includegraphics[width=\textwidth]{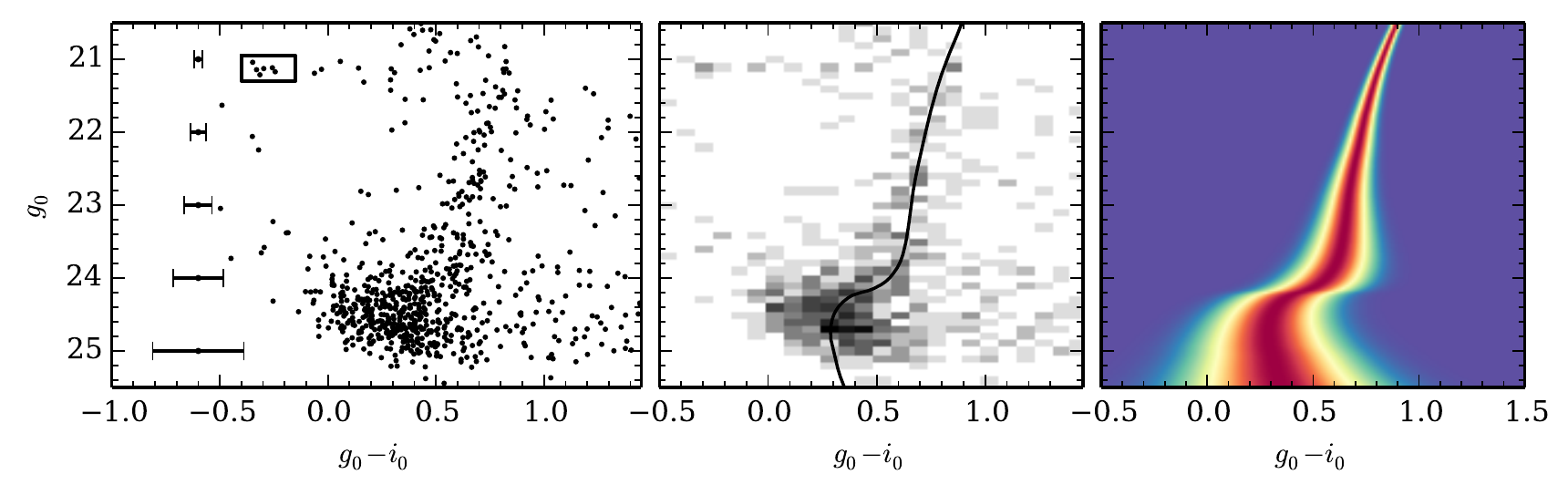}
\caption{Left: Colour-magnitude diagram of stars within the half-light radius of Hercules. Centre: Hess diagram of the same region as left, with 15 Gyr model isochrone from the Dartmouth Stellar Evolution Database \citep{2008ApJS..178...89D} overlaid. Right: Isochrone mask based on the model isochrone.  The mask provides weights to the Hercules catalog based on proximity; stars closer to the isochrone are weighted more heavily. Note that the tip of the red giant branch is not visible due to the saturation point of the data.}
\label{fig:mask}
\end{figure*}

To that end, the best-fitting model isochrone and known distance to Hercules were used to produce a mask, separating MW foreground stars from Hercules stars in the CMD.  Considering the old ($>\,$12 Gyr), metal poor ($\simeq\,-2.41$), population of Hercules, an isochrone of age 15\,Gyr,  [Fe/H]= -2.49\,dex and [$\alpha$/Fe]=0.6\,dex from the Dartmouth Stellar Evolution Database \citep{2008ApJS..178...89D} was used (see Figure \ref{fig:mask}).  Rather than applying a straight cut along the isochrone, a graduated mask was applied, comprised of weights varying with proximity to the isochrone.  This was done to make allowances for photometric uncertainties, and provide a more rigorous method of discrimination than a straight cut selection box.  Note that although a horizontal branch is clearly visible in the CMD in Figure \ref{fig:mask}, it is not included in the mask.  These stars are used as individual tracers of the Hercules population later in the analysis. The tip of the red giant branch (RGB) is not visible due to the saturation point of the data. However, based on the luminosity function of the model isochrone, there are relatively few member stars populating this part of the CMD and the extra signal this would provide against the foreground would be minimal.

In order to create the mask, the isochrone was binned into 0.02 mag increments in \textit{g}, and a Gaussian function was calculated along the colour range for each bin, with the width determined by the colour uncertainty, and an amplitude of 1.  This gave stars lying on the isochrone the highest weight.  The mask is shown in the right panel of Figure \ref{fig:mask}.  The shape of the mask reflects the change in photometric uncertainty with magnitude, and encompasses the stellar population seen within the half-light radius of Hercules.

Once complete, the mask was applied to all stars in the CMD, assigning each star a weight, $w$, between 0 and 1.  This immediately enabled the selection of the more likely Hercules stars, by employing a cutoff at any chosen weight.  By lowering the cutoff weight, the selection region about the isochrone could be widened, and similarly the region could be narrowed by choosing a more strict cutoff.  This provided a flexible means by which to analyse the Hercules population.

\section{Analysis}
\label{sec:analysis}
The search for over-densities was undertaken in two main steps: the detection and formal identification of potential over-densities, followed by a comparison of each detection to a control region to quantify its significance.  Sections \ref{sec:detection} and \ref{sec:significance} describe these procedures in detail.

\subsection{Detecting Over-Densities}
\label{sec:detection}
\begin{figure*}
\centering
\includegraphics[width=\textwidth]{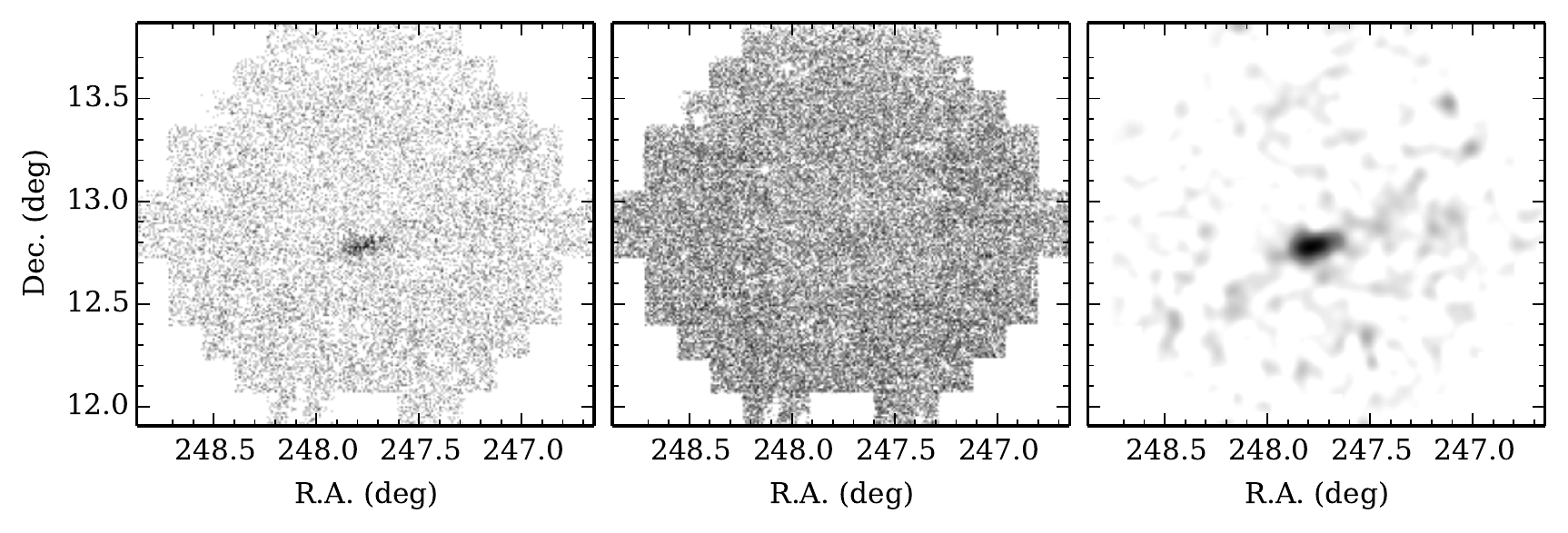}
\caption{Demonstration of the binning and smoothing process.  Left:  2D histogram of stars in the DECam field consistent with Hercules $w\geq0.2$.  Centre:  the complement of the left histogram ($w<0.2$) forming a `foreground map'.  Right: difference of the two histograms, smoothed with a Gaussian filter of 3 pixel ($120\arcsec$) kernel size.}
\label{fig:histogram}
\end{figure*}

In order to detect over-densities of Hercules stars in the DECam field, the distribution of stars in our catalogue was binned into a normalised 2D histogram of R.A. and DEC. Using $40\arcsec \times 40\arcsec$ bins, an image of 200$\times$177 pixels was created, where the pixel values reflected $count_{bin}/count_{sample}/area_{bin}$.  To maximise the contrast between the Galactic foreground and Hercules stars, two separate histograms were created (see Figure \ref{fig:histogram}) using the weights described in Section \ref{sec:discrimination}. The first contained only stars close to the model isochrone ($w \geq 0.2$), and the second contained everything else.  By subtracting this `foreground' map from the isochrone based histogram, artificial structures and foreground fluctuations were removed, leaving a well defined outline of Hercules and several potential over-densities (right panel in Figure \ref{fig:histogram}). This residual image was then convolved with a Gaussian filter of  varying kernel size (3, 5, and 7 pixels; smoothing over $120\arcsec, 200\arcsec$ and $280\arcsec$ respectively), in order to identify features on different scales.

The stellar features detected in this final smoothed histogram were formally identified and labelled using the Python package \textit{scipy.ndimage}\footnote{http://docs.scipy.org/doc/scipy/reference/ndimage.html}.  This package considers all pixels in an image above a specified threshold, and defines individual `segments' based on adjoining pixels.  Each segment was given a label for identification and later analysis.

We performed some experimentation in order to select a suitable threshold for the analysis.  We calculated the mean and standard deviation of the pixel values in the smoothed histogram, and applied threshold values as some multiple of this standard deviation ($\sigma_t$).  If the threshold was set too low, it was found that the algorithm returned many detections that were subsequently determined not to be significant; conversely, if the threshold was set too high then the algorithm found only the central body of Hercules. Thresholds of $1\sigma_t$ and $1.5\sigma_t$ struck an acceptable balance between these two extremes, and we performed the segmentation process twice with these two different thresholds in order to probe the extent of each detected over-density. Pixels with values $> 4 \sigma_t$ above the mean were considered part of the main body of Hercules and excluded from labelling (this region was slightly larger than the area contained inside the  half-light radius).  We also performed the segmentation process using a threshold of $0.5\sigma_t$ in order to explore the very low-level structure of the central over-density surrounding the main body of Hercules. Although we do not report the results at this threshold in the same detail as for the other two thresholds (in order to maintain the clarity of the paper), we return to this calculation in Section 5.

\begin{figure}
\centering
\includegraphics[width=8cm]{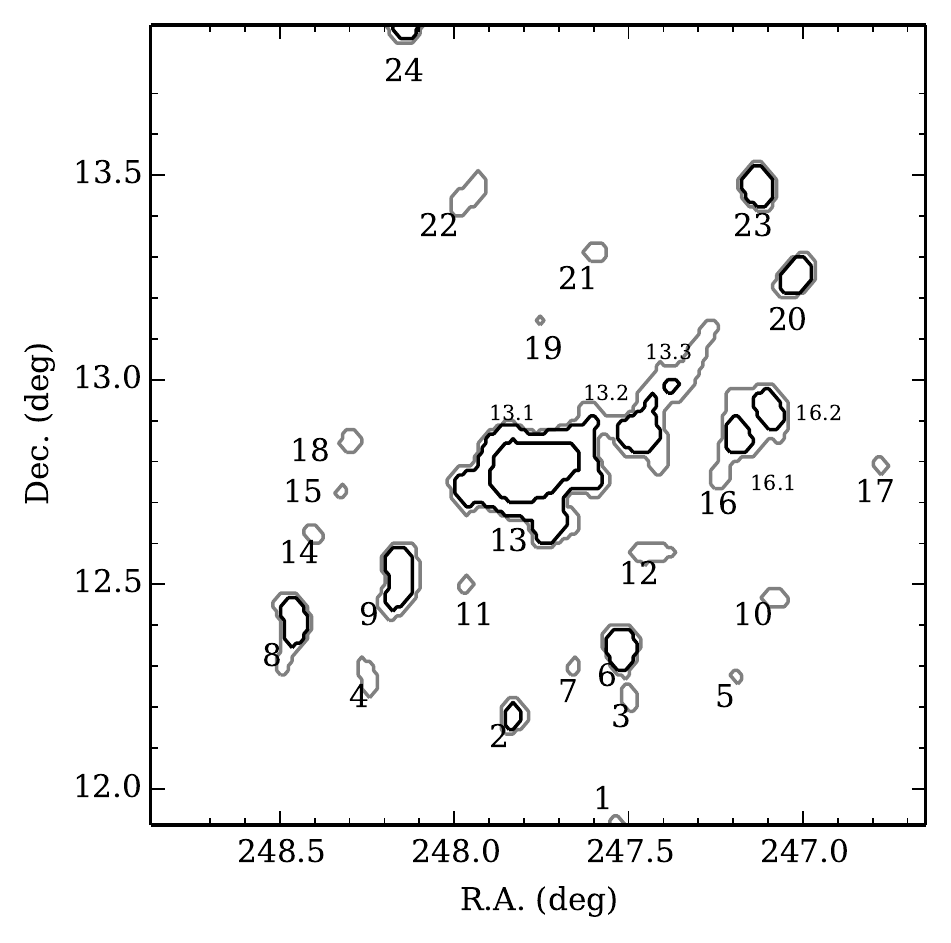}
\caption{Labelling of segments from the segmentation process for the 3 pixel ($120\arcsec$) kernel smoothed image.  $1\sigma_t$ threshold detections are in grey, and $1.5\sigma_t$ threshold detections are in black.  The labelling scheme is such that the label of the $1.5\sigma_t$ threshold detections take on the label of their $1\sigma_t$ threshold counter-parts. Where there is more than one corresponding segment, labelling takes a decimal form (e.g. Segments 13.1, 13.2, and 13.3 are the  $1.5\sigma_t$ threshold counter-parts to segment 13).  Note that the very centre of Hercules is excluded from segmentation. }
\label{fig:segments}
\end{figure}

Using this regime, individual features were identified as `segments' and labelled. As the kernel increased from $120\arcsec$ to $200\arcsec$ and $280\arcsec$, the smaller details disappeared and only the larger features remained apparent.  As a check, this process was also performed with a $60\arcsec$ kernel, at which point the level of detection `noise' became too large.  The results from the $120\arcsec$ kernel are therefore presented here as it represents the best compromise between detail and noise.  Figure \ref{fig:segments} shows the over-densities for the two thresholds, smoothed with the $120\arcsec$ kernel.  Note that the labelling based on the $1 \sigma_t$ threshold was adopted for the $1.5 \sigma_t$ threshold, in order to compare similar over-densities (e.g. OD 13 surrounding the main body of Hercules at the 1$\sigma_t$ threshold, decomposes into three segments at $1.5\sigma_t$, labelled OD 13.1, OD 13.2, and OD 13.3 respectively).

During the segmentation process, a control region was also defined for use in later analysis, consisting of all pixels in the 2D-histogram with a value less than the detection threshold.  This ensured that no over-density was used in the control sample, and the sample was evenly distributed across the DECam field.

Once the identification and labelling of segments were complete, the stars which fell inside the segment boundaries were extracted from the catalogue, regardless of their isochrone weight.  This ensured that the significance of each feature was analysed indiscriminately.  Similarly, all stars corresponding to the control region were extracted, forming a sample for significance testing.

\subsection{Significance of Detections}
\label{sec:significance}

Once the stars of each over-density had been extracted, the significance of each feature was tested relative to the model isochrone.  As well as this, a brief test was conducted to check if extinction or variations in completeness across the field could be responsible for the over-densities detected.  The location of the over-densities was compared to the extinction map from section \ref{sec:photom} to check for correlations.  It was found that the peak-to-peak amplitude of the extinction had to be increased by a factor of three before there were even hints of a correlation.  As this equates to approximately five times the variation in the derived 50\% completeness level across the field, it was deemed that these factors do not play a significant role in our detection of over-densities.

\begin{figure}
\centering
\includegraphics[width=8cm]{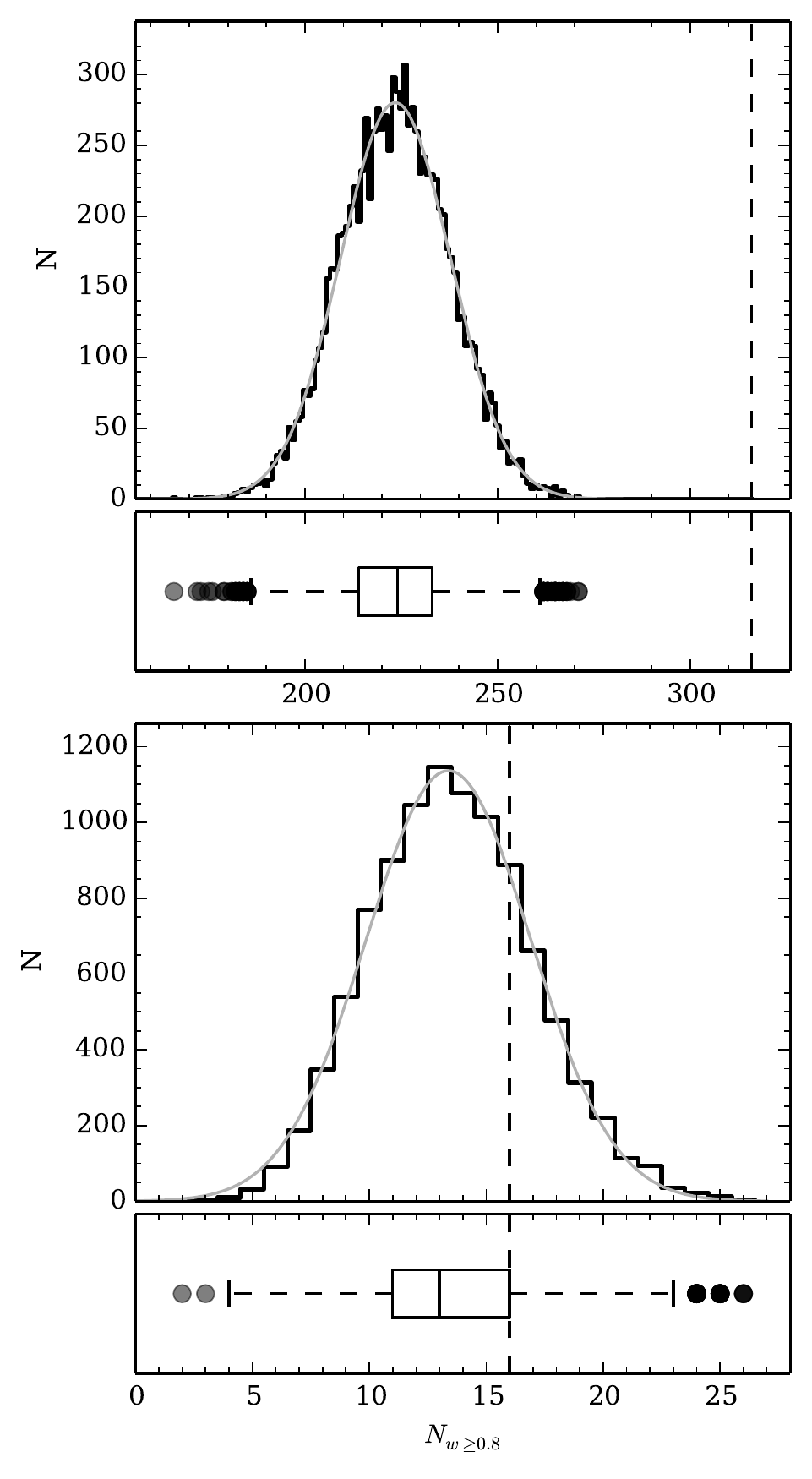}
\caption{Results of the control sampling process for two over-density detections (Top: OD 13, Bottom: OD 22).  The histograms represent the frequency distribution of $N_{w\geq0.8}$ values from the 10,000 control sub-samples taken for each over-density.  The black dashed lines represents $N_{w\geq0.8}$ for each over-density. The solid grey lines show a Gaussian fit to each frequency distribution. The box plot represents the median and 25th/75th percentiles of the distribution, with the whiskers showing the most extreme data points not considered outliers, and the filled circles representing the outliers.  Note that the outliers are shown at 50\% opacity so underlying points may be seen.}
\label{fig:bwplots}
\end{figure}

A feature was determined to be significant if a notable number of its stars are close to the isochrone in the CMD, i.e. $w \geq 0.8$.  The number of stars with $w\geq0.8$ ($N_{w\geq0.8}$) was counted and recorded for each segment.  The value of $N_{w\geq0.8}$ for each detection was then tested against the control sample defined in Section \ref{sec:detection}.  In order to carry out this test, a sub-sample of stars from the control sample was drawn randomly such that the number of stars in the sub-sample was equal to the number of stars in a given segment.   $N_{w\geq0.8}$  was counted and recorded for this control sub-sample, similarly to the segment.  This process was repeated 10,000 times.  The resulting frequency distribution was then compared to the value recorded for the segment.  Figure \ref{fig:bwplots} illustrates the results for two of  the over-densities detected (OD 13 and OD 22). The frequency distribution of the control sampling is displayed as a histogram, with a box and whisker plot showing the range of possible values the control samples could take. The value of $N_{w\geq0.8}$ recorded for the segment is marked with a black dashed line.  The result for OD 13 (top) represents a significant detection, where $N_{w\geq0.8}$ for the over-density is greater than the range of values found for the control samples.  The result for OD 22 (bottom), on the other hand, suggests that this over-density could be explained as a random fluctuation in the foreground population, with $N_{w\geq0.8}$ for the over-density being quite close to the mean control sample value.

A `significance value' ($\zeta$) was determined for each test result, as a means of identifying the most significant detections, and comparing these results to the null hypothesis.  The value of $\zeta$ was generated by calculating how far  $N_{w\geq0.8}$  was away from the mean of the control sample ($\langle N_{w\geq0.8}(CS)\rangle$).  This was achieved by fitting a Gaussian to the frequency distribution of $N_{w\geq0.8}$ values for the control samples (see Figure \ref{fig:bwplots}), and expressing the distance in units of the standard deviation ($\sigma$):

\begin{displaymath}
\zeta = \frac{N_{w\geq0.8}(OD) - \langle N_{w\geq0.8}(CS)\rangle}{\sigma}
\end{displaymath}

\begin{figure*}
\centering
\includegraphics[width=\textwidth]{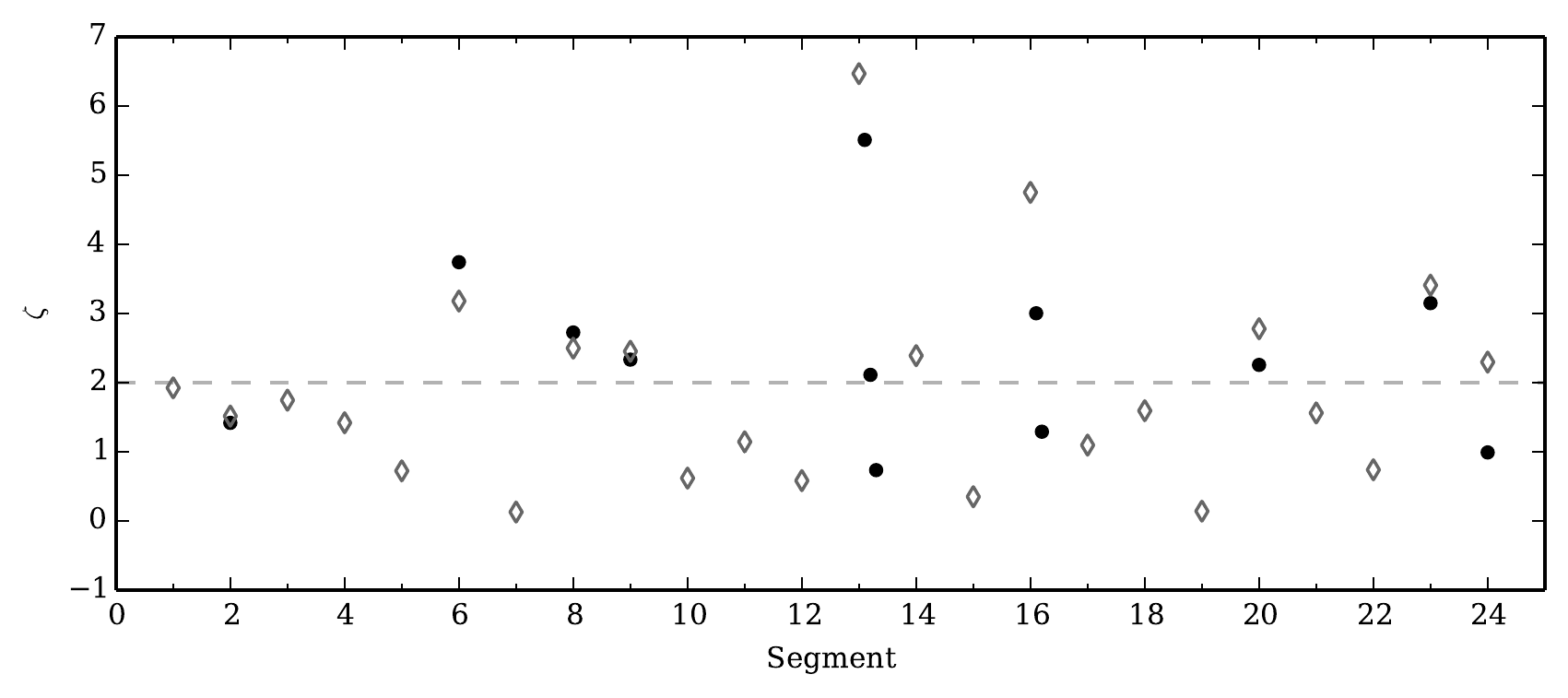}
\caption{The significance value, $\zeta$, for each of the over-densities detected (corresponding to Tables \ref{tab:results1a} and  \ref{tab:results1b}).  Grey diamonds denote a 1$\sigma_t$ threshold detection, and black circles denote a 1.5$\sigma_t$ detection (corresponding to Figure \ref{fig:segments}). }
\label{fig:results}
\end{figure*}

A detection was considered significant if $\zeta\geq2.0$.  The results of this analysis are summarised in Tables \ref{tab:results1a} and \ref{tab:results1b}, and presented graphically in Figure \ref{fig:results}.  Diamonds and circles in this figure represent detections based on the $1\sigma_t$ and $1.5\sigma_t$ thresholds respectively.  There are nine over-densities that are significant.  Several of these are sufficiently large that they appear significant in the larger 5 pixel and 7 pixel convolutions. These will be discussed in more detail later.

\begin{table}
\begin{center}
\caption{Significance values for the $1\sigma_t$ detected Hercules over-densities, where segment labelling corresponds to the grey outlines in Figure \ref{fig:segments}.}
\label{tab:results1a}
\begin{tabular}{ccrrrr}
\tableline\tableline
Segment && $N_{*}$ & $N_{w\geq0.8}(OD)$ & $\langle N_{w\geq0.8}(CS)\rangle$ & $\zeta$ \\
\cline{1-1}\cline{3-6}
OD 1 && 18  & 4  & 1  & 1.93\\
OD 2 && 136  & 17  & 12  & 1.52\\
OD 3 && 65  & 10  & 6  & 1.75 \\
OD 4 && 74  & 10  & 7  & 1.42 \\
OD 5 && 16  & 2  & 1  & 0.73   \\
OD 6 && 309  & 44  & 28  & 3.18   \\
OD 7 && 21  & 1  & 2  & 0.13  \\
OD 8 && 377  & 48  & 34  & 2.50  \\
OD 9 && 530  & 64  & 48  & 2.45  \\
OD 10 && 55  & 6  & 5  & 0.62 \\
OD 11 && 19  & 3  & 2  & 1.15\\
OD 12 && 89  & 9  & 8  & 0.59 \\
OD 13 && 2468  & 316  & 224  & 6.47 \\
OD 14 && 39  & 8  & 3  & 2.39 \\
OD 15 && 4  & 0  & 0  & 0.35   \\
OD 16 && 908  & 123  & 82  & 4.75   \\
OD 17 && 20  & 3  & 2  & 1.10 \\
OD 18 && 53  & 8  & 5  & 1.60\\
OD 19 && 5  & 0  & 0  & 0.14  \\
OD 20 && 257  & 36  & 23  & 2.78 \\
OD 21 && 53  & 8  & 5  & 1.56 \\
OD 22 && 150  & 16  & 13  & 0.74\\
OD 23 && 226  & 35  & 20  & 3.41\\
OD 24 && 21  & 5  & 2  & 2.30 \\

\tableline \tableline
\end{tabular}
\end{center}
\end{table}

\begin{table}
\begin{center}
\caption{Significance values for the $1.5\sigma_t$ detected Hercules over-densities, where segment labelling corresponds to the black outlines in Figure \ref{fig:segments}.}
\label{tab:results1b}
\begin{tabular}{ccrrrr}
\tableline\tableline
Segment && $N_{*}$ & $N_{w\geq0.8}(OD)$ & $\langle N_{w\geq0.8}(CS)\rangle$ & $\zeta$ \\
\cline{1-1}\cline{3-6}
OD 6 && 207  & 34  & 19  & 3.74\\
OD 8 && 183  & 27  & 17  & 2.73\\
OD 9 &&  282  & 37  & 26  & 2.34\\
OD 13.1&&1019  & 144  & 93  & 5.51 \\
OD 13.2 &&255  & 33  & 23  & 2.11 \\
OD 13.3 && 16  & 2  & 1  & 0.74 \\
OD 16.1 &&104  & 18  & 9  & 3.00 \\
OD 16.2 &&168  & 20  & 15  & 1.29 \\
OD 20&& 170  & 24  & 16  & 2.26\\
OD 23&& 144  & 24  & 13  & 3.15 \\
OD 24&& 13  & 2  & 1  & 0.99 \\
\tableline \tableline
\end{tabular}
\end{center}
\end{table}
  
\begin{figure}
\centering
\includegraphics[width=8cm]{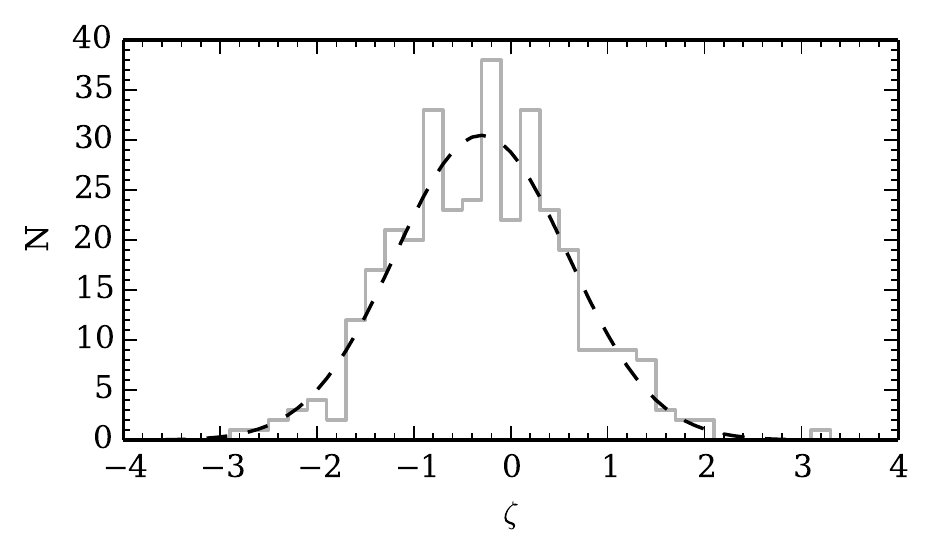}
\caption{Histogram of possible values of $\zeta$ for 300 blind segments taken from regions of the Hercules field devoid of over-density detections.}
\label{fig:nullresults}
\end{figure}

In order to test the null hypothesis, the significance analysis, as described in section \ref{sec:significance}, was repeated for 300 randomly selected segments from the control region, varying in size between 4 - 18 sqr arcmin.  Figure \ref{fig:nullresults} shows the resulting frequency distribution of $\zeta$ values for the 300 segments.  The mean and standard deviation are computed as -0.30 and 0.89, respectively, further supporting the significance of detections with $\zeta\geq2.0$.

\begin{figure}
\centering
\includegraphics[width=8cm]{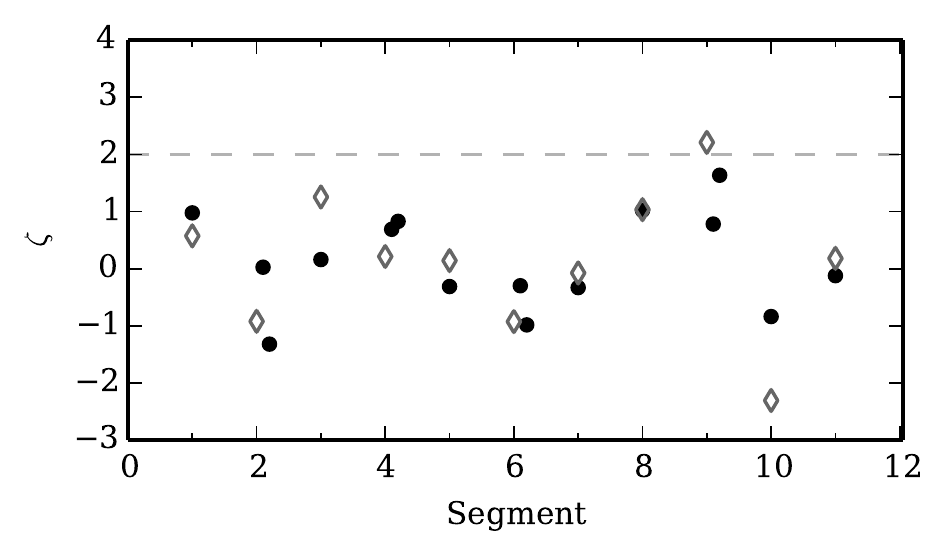}
\caption{Similar to Figure \ref{fig:results}, this plot shows the results for segmentation based on the nuggets from \citet{2009ApJ...704..898S}.  Black points represent the 3$\arcmin$ segments, and grey points represent the 6$\arcmin$ segments. }
\label{fig:sandresults}
\end{figure}

The significance test was also performed on the `nuggets' reported by \citet{2009ApJ...704..898S} (see their Figure 9).  The majority of these nuggets appear quite small, varying from less than an arcminute to a few arcminutes in size. Using the coordinates of the nuggets from this figure as centre, two different sized boxes of 3$\arcmin$ and 6$\arcmin$ width were defined for each nugget in order to examine their extent, and ensure a large enough sample of stars in the nuggets for our statistical analysis.  The segmentation and significance testing was performed for each box size, mimicking the process used for detecting our over-densities.  The results are summarised in Tables \ref{tab:sanda} and \ref{tab:sandb}, and Figure \ref{fig:sandresults}.  Only one of these nuggets coincides with an over-density in our list: NGT 9 (NGT 9.2 at $3\arcmin$ width) showed $\zeta=2.21$, while our corresponding over-density (OD 13.2) showed $\zeta=2.11$.  All other nugget segments have significance values considerably below our $\zeta=2$ limit. Since their data was somewhat deeper than ours, it is unsurprising that we may not detect all of the nuggets they did.  It should be noted that with the exception of OD 13, there is little overlap between our over-densities and the fields observed by \citet{2009ApJ...704..898S}.  Several of our less significant detections ($\zeta<2.0$) have very small amounts of overlap with these fields.  Since these detections are not strong, it is unlikely that partial coverage  would yield a detection. There is however, partial over-lap with OD 9 and OD 16, which we find to be significant.  These over-densities appear less significant at the $1.5\sigma_t$ threshold, indicating they are extremely diffuse.  Partial coverage of such diffuse over-densities may cause them to go unnoticed, even in a deeper data set.  With regard to the main halo-like body of OD 13, there is remarkable correlation with the contours in Figure 9 from \citet{2009ApJ...704..898S} (upper central panel).

\begin{table}
\begin{center}
\caption{Significance values for `nuggets' depicted by \citet{2009ApJ...704..898S}, using a $6\arcmin\times6\arcmin$ box for segmentation. The segments are shown in Figure \ref{fig:summary}.}
\label{tab:sanda}
\begin{tabular}{ccrrrr}
\tableline\tableline
Segment && $N_{*}$ & $N_{w\geq0.8}(NGT)$ & $\langle N_{w\geq0.8}(CS)\rangle$ & $\zeta$\\
\cline{1-1}\cline{3-6}
NGT 1 && 321  & 32  & 29  & 0.57 \\
NGT 2 && 1022  & 84  & 93  & -0.92\\
NGT 3 && 248  & 28  & 22  & 1.25   \\
NGT 4 && 570  & 53  & 52  & 0.21  \\
NGT 5 && 248  & 23  & 22  & 0.14  \\
NGT 6 && 651  & 52  & 59  & -0.93 \\
NGT 7 && 260  & 23  & 23  & -0.07 \\ 
NGT 8 && 278  & 30  & 25  & 1.03 \\
NGT 9 && 717  & 82  & 65  & 2.21\\
NGT 10 && 308  & 16  & 28  & -2.31\\ 
NGT 11 && 291  & 27  & 26  & 0.18  \\

\tableline \tableline
\end{tabular}
\end{center}
\end{table}

\begin{table}
\begin{center}
\caption{Significance values for `nuggets' depicted by \citet{2009ApJ...704..898S}, using a $3\arcmin\times3\arcmin$ box for segmentation.The segments are shown in Figure \ref{fig:summary}.}
\label{tab:sandb}
\begin{tabular}{ccrrrr}
\tableline\tableline
Segment && $N_{*}$ & $N_{w\geq0.8}(NGT)$ & $\langle N_{w\geq0.8}(CS)\rangle$ & $\zeta$\\
\cline{1-1}\cline{3-6}
NGT 1&& 94  & 11  & 8  & 0.98 \\
NGT 2.1 && 353  & 32  & 32  & 0.03 \\
NGT 2.2 && 68  & 2  & 6  & -1.32  \\
NGT 3&& 87  & 8  & 8  & 0.16\\
NGT 4.1&& 74  & 8  & 7  & 0.69 \\
NGT 4.2&& 107  & 12  & 10  & 0.83 \\
NGT 5 && 84  & 6  & 7  & -0.31  \\
NGT 6.1 && 84  & 6  & 7  & -0.30 \\
NGT 6.2 && 164  & 11  & 15  & -0.98 \\
NGT 7&& 101  & 8  & 9  & -0.33 \\
NGT 8&& 84  & 10  & 7  & 1.02 \\
NGT 9.1&& 189  & 20  & 17  & 0.78 \\
NGT 9.2&& 95  & 13  & 8  & 1.64 \\
NGT 10&& 106  & 7  & 10  & -0.84 \\
NGT 11&& 95  & 8  & 8  & -0.12 \\

\tableline \tableline
\end{tabular}
\end{center}
\end{table}

\section{Results}
\label{sec:results}

Further to the investigation of the spatial distribution of the over-densities detected, we tried to estimate the luminosity of these new over-densities.  For that purpose, the over-densities were subdivided into four regions.  The first region was Hercules itself, as defined during the segmentation process.  It covers about 1.75 times the area of the half-light ellipse and was excluded from our analysis.  The second was the over-density surrounding Hercules,  OD 13.  The third was all other over-densities with $\zeta\geq3.0$, and the fourth region included those over-densities with $2.0\leq\zeta<3.0$.  For each region, the combined flux of all stars with $w\geq0.8$ was calculated.  An estimate of the foreground flux was made by summing the flux from stars with $w\geq0.8$ for fields drawn randomly from the control region, equal in area to each of the over-density regions.  This was performed 1000 times for each of the four over-density regions, in order to build up a frequency distribution for which a statistically robust mean and standard deviation could be determined.  The mean value for each was taken to represent the foreground over that area of sky, and subtracted from the flux counted for the corresponding over-density region.  The remaining flux was considered to represent an upper limit for the luminosity of the over-density regions.  The extended region 2, surrounding the main body of Hercules (region 1), was calculated to have $52\pm12\%$ of the flux of region 1, spread over approximately 2.7 times the area.  Similarly, Region 3 contains $42\pm9\%$ as much flux as region 1, spread over approximately 1.4 times the area.  Region 4 contains $25\pm8\%$ as much flux as region 1 spread over approximately 1.2 times the area.  In total, the combined flux found in all significant over-densities (regions 2, 3, and 4) is of the same order ($119\pm29\%$) as the flux of Hercules itself, thus a significant portion of Hercules' stellar population has been tidally disrupted.

\begin{figure*}
\centering
\includegraphics[width=\textwidth]{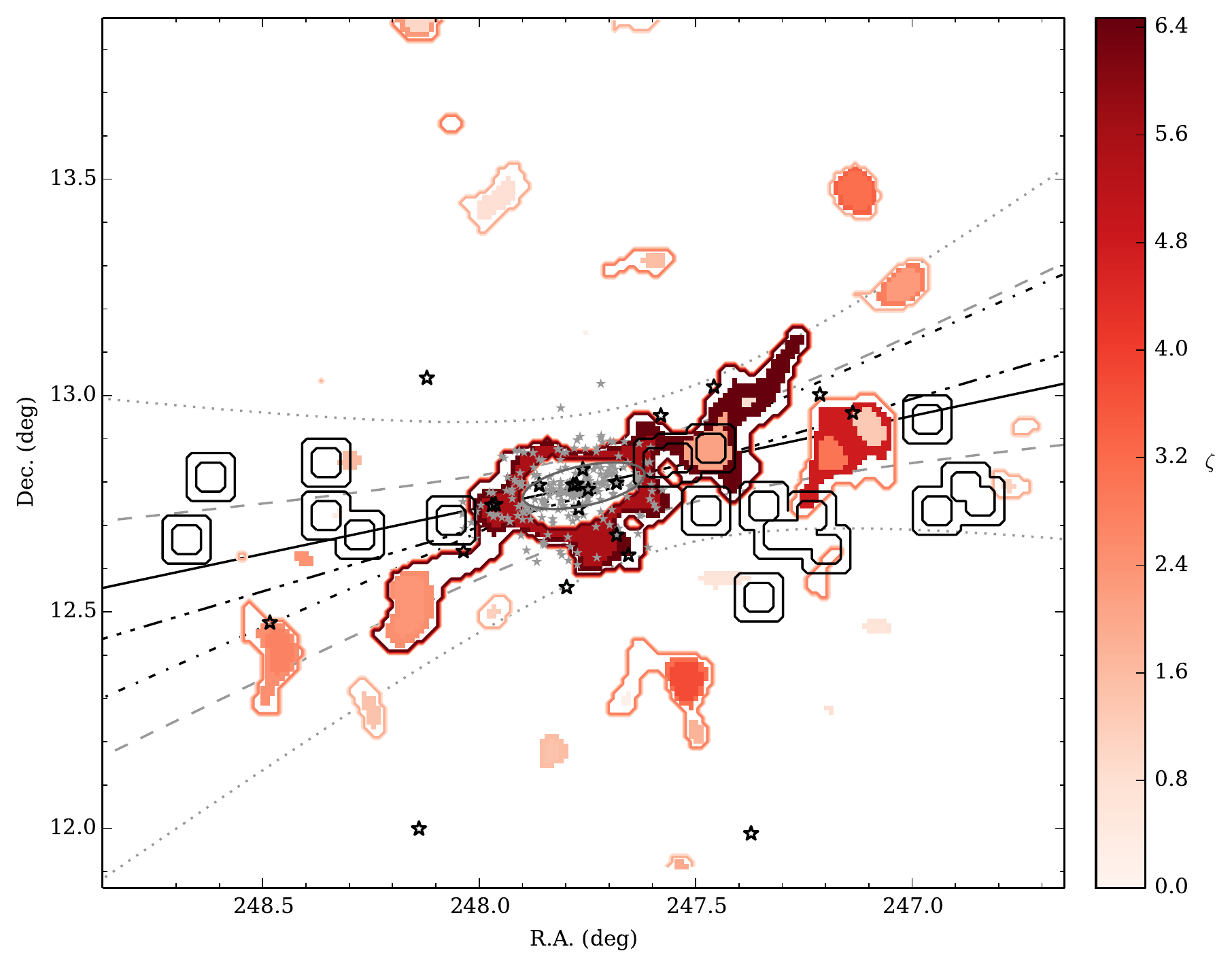}
\caption{Summary figure displaying compiled results from this analysis.  The significance of $1\sigma_t$ and $1.5\sigma_t$ detections are colour coded according to $\zeta$ ($1.5\sigma_t$ detections appear `inside' the $1\sigma_t$ detections). Coloured contours reflect those over-densities detected at the $0.5\sigma_t$ threshold with $\zeta\geq2.0$.  Our segmentation of the nuggets from \citet{2009ApJ...704..898S} is shown as an overlay of black outlined boxes.  Candidate blue horizontal branch stars, selected from the CMD (see Figure \ref{fig:mask}), are shown as open black stars.  The elongation of Hercules and the orbital path proposed by \citet{2010ApJ...721.1333M} are demonstrated by a black solid line and grey ellipse.  Small grey-filled stars represent stars identified as potential Hercules members, through relative proper motions, from Table 2 of \citet{2014A&A...570A..61F}. The dash-dot line represents the inclination of Hercules in this data set based on a maximum likelihood bivariate normal fit to stars inside the central 9\arcmin$\times$9\arcmin. The dash-dot-dot line represents a weighted least squares fit to the position of the over-densities, using $\zeta$ as weights. Confidence intervals for this fit at the 68 and 95 percentile limits are shown by the dashed and dotted grey lines respectively.}
\label{fig:summary}
\end{figure*}

The summary of our analysis is illustrated in Figure \ref{fig:summary}. The new over-densities are colour-coded according to their significance, with the $1.5\sigma_t$ detections laid over the $1\sigma_t$ detections.  We have also over-laid the outline of those $0.5\sigma_t$ detections with $\zeta>2.0$, in order to provide the broadest illustration of Hercules.  The segmentation of the nuggets from \citet{2009ApJ...704..898S} is shown as an overlay of black outlined boxes.  Stars consistent with the blue horizontal branch (BHB) of Hercules ($21.0<g_0<21.2$ $-0.40<(g-i)_0<-0.15$), are shown as open black stars.  The elongation of Hercules and the postulated orbital path from \citet{2010ApJ...721.1333M} are demonstrated by a grey ellipse and black solid line respectively.  We note that the orbital path presented by  \citet{2010ApJ...721.1333M} represents a viable scenario in which Hercules has had a close encounter with the MW, and the orbital path was assumed to be aligned with the direction of elongation of Hercules.  This is consistent with the idea that Hercules has very recently passed peri-galacticon, not yet having flipped its tidal tails in an orientation toward the MW \citep{2009MNRAS.400.2162K}.  Small grey-filled stars represent stars identified as potential Hercules members, by means of relative proper motions, from Table 2 of \citet{2014A&A...570A..61F}.  This table includes all 528 of the initial proper motion candidates identified by \citet{2014A&A...570A..61F}, which they state contains a level of remaining field contamination. In order to reduce the number of contaminants in the sample, we preferentially selected stars sitting closest to the  red-giant and horizontal branches in the CMD. The distribution of these 239 proper motion, CMD, selected stars shows a remarkable correlation with the outline of the extended body of Hercules.

The overall distribution of the over-densities detected appears roughly aligned with the major axis of Hercules.  However, two of the significant detections (OD 6 and OD 23) lie at notably different orientation angles to the major axis.  These over-densities could be free-floating tidal debris following Hercules on slightly different orbits. As a means of further investigation,  the astroML\footnote{http://www.astroml.org/} \citep{astroML} machine learning package was employed to determine the ellipticity and angular orientation of Hercules in our data set.  Taking the central 9\arcmin$\times$9\arcmin (containing the half-light radius of Hercules), a maximum likelihood bivariate normal fit was performed.  An ellipticity of $\epsilon=0.63\pm0.02$ was found, which is in agreement with the published value of $0.68\pm0.08$ \citep{McConnachie:2012fh}.  An orientation angle of $113.8\pm0.6^\circ$ was found, which is slightly steeper than the published value of $102^\circ\pm4^\circ$\citep{McConnachie:2012fh}.  In the scenario from \citet{McConnachie:2012fh}, where the orbital path is aligned with the position angle of the major axis of Hercules, this would indicate a steeper orbital path as plotted with the dash-dot line in Figure \ref{fig:summary}. This path lies closer to several of the leading and trailing over-densities.  To test the correlation between the properties of the main stellar body of Hercules and the distribution of the newly identified stellar over-densities, we fit a line to the positions of all of the over-densities, using the $\zeta$ values as weights for the fit.  This line is shown in Figure \ref{fig:summary} as the dash-dot-dot line.  Confidence bands for this fit at the $68\%$ and $95\%$ level are shown by the dashed and dotted grey lines respectively.  Both the orbital path from \citet{2010ApJ...721.1333M}, and our revised orientation of the major axis of Hercules, fall inside the $68\%$ confidence band.

A clear trend is also detected in the distribution of the BHB star candidates, following the elongation of Hercules.  Since these are BHB star candidates (as photometrically defined by the black box in the CMD in Figure \ref{fig:mask}), the expectation is that they belong to Hercules, and therefore provide a good tracer of its stellar population.  Tests were performed to see if the distribution of BHB stars could be reproduced with a random selection of stars from across the field.  There are 21 BHB star candidates, however, the seven stars within the half-light radius were excluded from testing due to the high probability of membership.  Thus, 14 stars remained in the outer regions.  Of these, six appear to coincide with (or sit on top of) the detected over-densities; in addition, six are observed to sit within the 68\% confidence band and 11 fall within the 95\% band.  It was found that, drawing a random sample of 14 stars from anywhere in the field, six or more coincided with an over-density 1364 times out of $10^7$ trials (0.01\%).  A similar test was performed to assess the likelihood of stars falling inside the $68\%$ and $95\%$ confidence bands.  It was found that six or more out of 14 stars lie inside the $68\%$ interval 0.09\% of the time (out of $10^7$ trials).  Similarly, 11 or more out of 14 stars fall inside the $95\%$ confidence band 2\% of the time (out of $10^7$ trials).  This shows that the distribution of BHB candidates we detect beyond Hercules' radius is in agreement with the alignment of the substructures at a confidence level greater than 98\%.

\section{Summary}
\label{sec:summary}
The tidal disruption of the Hercules dwarf galaxy has been alluded to for some time \citep{2007ApJ...668L..43C, 2009ApJ...706L.150A, 2009ApJ...704..898S, 2010ApJ...721.1333M}, however the full extent to which it occurs was not easy to quantify due to technical limitations.  Our deep, photometric analysis of 3 sqr deg around Hercules, with the Dark Energy Camera, reveals that this dwarf galaxy has undergone significant tidal disruption, with nine statistically significant over-densities extending to more than $2\,$kpc from its centre. Most over-densities are distributed along the major axis, with all but two falling inside the 95\% confidence band of the likely orbital path.  We estimated the combined luminosity of these over-densities relative to the Hercules galaxy itself.  The total number of stars in these over-densities is of the same order as the total number of stars in the main body of Hercules, highlighting the severe tidal disruption this dwarf galaxy has experienced in the tidal field of the Milky Way.

Although we find little correlation with the nuggets reported by \citet{2009ApJ...704..898S} (with the exception of the shape of the extended region surrounding the main body of Hercules), we find excellent correlation with the proper-motion identified stars from Table 2 of \citet{2014A&A...570A..61F}.  Given the distinctly different method used by \citet{2014A&A...570A..61F}, the correlation of these results with our own provides extra confidence in the position of our over-densities, and encourages the use of relative proper motion as a means of identifying membership.

Our results provide useful new observational constraints for theoretical modelling. Recent work by  \citet{2015MNRAS.446..144B} raises questions about the current picture of the orbit of Hercules, and suggests that a path aligned along the major axis of the dwarf may not necessarily be correct. Our results are consistent with the orbital path proposed by \citet{2010ApJ...721.1333M}, although both the position angle of the major axis, and the alignment of the substructures suggest a slightly steeper orientation.  Interestingly, the position angle of the major axis that we determine does not line up perfectly with the alignment of the substructures (although the two are consistent within the 68\% confidence band). This suggests a picture similar to that for the globular cluster Palomar 5, where the orbital path is not quite aligned with the leading and trailing tidal arms  \citep{2004AJ....127.2753D,2009AJ....137.3378O}. It may also be an indication that the tidal tails are transitioning to an orientation aligned towards the Milky Way as the dwarf approaches apo-galacticon \citep{2009MNRAS.400.2162K}. Finally, two of the significant over-densities we detect sit outside the confidence bands of the substructure alignment. This may suggest that Hercules has made more than one peri-galactic pass, and is accompanied by free-floating debris. This would be consistent with infall timescales of ~2-5 Gyr found in the literature \citep{2012MNRAS.425..231R,2015MNRAS.446..144B}, and the destructive peri-galactic passage described by  \citet{2010ApJ...721.1333M}.

The identification of statistically significant over-densities provides new opportunities for spectroscopic follow up and further analysis in the ELT era, as well as new data that can be added to test and calibrate theoretical models.

\acknowledgments
The authors wish to thank David James for his assistance during the DECam observing run, and Agris Kalnajs for discussions during the analysis of this data.  The authors also thank the referee for thoughtful discussion on the original manuscript. TAR acknowledges financial support through an Australian Postgraduate Award. HJ and GDC acknowledge the support of the Australian Research Council (ARC) through Discovery Projects DP120100475 and DP 150100862.  ADM is grateful for support from the Australian Research Council (ARC) in the form of an Australian Research Fellowship (Discovery Projects Grant DP1093431). This project used data obtained with the Dark Energy Camera (DECam), which was constructed by the Dark Energy Survey (DES) collaborating institutions: Argonne National Lab, University of California Santa Cruz, University of Cambridge, Centro de Investigaciones Energeticas, Medioambientales y Tecnologicas-Madrid, University of Chicago, University College London, DES-Brazil consortium, University of Edinburgh, ETH-Zurich, University of Illinois at Urbana-Champaign, Institut de Ciencies de l'Espai, Institut de Fisica d'Altes Energies, Lawrence Berkeley National Lab, Ludwig-Maximilians Universitat, University of Michigan, National Optical Astronomy Observatory, University of Nottingham, Ohio State University, University of Pennsylvania, University of Portsmouth, SLAC National Lab, Stanford University, University of Sussex, and Texas A\&M University. Funding for DES, including DECam, has been provided by the U.S. Department of Energy, National Science Foundation, Ministry of Education and Science (Spain), Science and Technology Facilities Council (UK), Higher Education Funding Council (England), National Center for Supercomputing Applications, Kavli Institute for Cosmological Physics, Financiadora de Estudos e Projetos, Funda‹o Carlos Chagas Filho de Amparo a Pesquisa, Conselho Nacional de Desenvolvimento Cient'fico e Tecnol—gico and the MinistŽrio da Cincia e Tecnologia (Brazil), the German Research Foundation-sponsored cluster of excellence ``Origin and Structure of the Universe" and the DES collaborating institutions.

\bibliographystyle{apj}
\bibliography{Hercules}

\end{document}